\documentclass[11pt,a4paper]{article}
\usepackage[utf8]{inputenc}
\usepackage[T1]{fontenc}
\usepackage{lmodern}
\usepackage{microtype}
\usepackage[margin=2.5cm]{geometry}
\usepackage{setspace}
\usepackage{parskip}
\usepackage{graphicx}
\usepackage{booktabs}
\usepackage{enumitem}
\usepackage{hyperref}
\hypersetup{colorlinks=true,linkcolor=blue,citecolor=blue,urlcolor=blue}
\usepackage{titling}
\usepackage{fancyhdr}
\usepackage{caption}
\usepackage{amsmath,amssymb}

\usepackage[
  backend=biber,
  style=numeric,
  sorting=none,
  giveninits=true,
  maxnames=1
]{biblatex}

\defbibheading{bibliography}[\refname]{\section*{#1}\vspace{-8pt}}

\renewbibmacro*{in:}{}

\AtEveryBibitem{%
  \clearfield{title}%
  \clearfield{subtitle}%
  \clearfield{number}%
  \clearfield{issue}%
  \clearfield{month}%
  \clearfield{day}%
  \clearfield{doi}%
  \clearfield{url}%
  \clearfield{isbn}%
  \clearfield{issn}%
  \clearfield{note}%
  \clearfield{addendum}%
  \clearfield{howpublished}%
  \clearfield{eprintclass}%
}

\renewbibmacro*{date}{\printfield{year}}

\DeclareFieldFormat{eprint}{arXiv:#1}

\DeclareBibliographyDriver{article}{%
  \printnames{author}\setunit{\addspace}%
  \printfield{year}\setunit{\addcomma\space}%
  \printfield{journaltitle}\setunit{\addcomma\space}%
  \printfield{volume}\setunit{\addcomma\space}%
  \iffieldundef{journaltitle}
    {\iffieldundef{eprint}{}{%
       \setunit{\addcomma\space}\printfield{eprint}}}
    {}%
  \finentry
}

\DeclareBibliographyDriver{online}{%
  \printnames{author}\setunit{\addspace}%
  \printfield{year}%
  \iffieldundef{eprint}{}{%
    \setunit{\addcomma\space}\printfield{eprint}}%
  \finentry
}
\DeclareBibliographyDriver{unpublished}{%
  \usebibmacro{bibindex}\usebibmacro{begentry}%
  \usebibmacro{author/translator+others}\setunit{\addspace}%
  \printfield{year}%
  \iffieldundef{eprint}{}{\setunit{\addcomma\space}\printfield{eprint}}%
  \usebibmacro{finentry}%
}

\defbibenvironment{bibliography}
  {}
  {}
  {\addspace
   \printtext[labelnumberwidth]{%
     \printfield{prefixnumber}%
     \printfield{labelnumber}}%
   \addhighpenspace}

\pretitle{\vspace*{-1cm}\begin{center}\LARGE\bfseries}
\posttitle{\par\end{center}\vskip 0.5em}
\preauthor{\begin{center}\large}
\postauthor{\end{center}}
\predate{\begin{center}\large}
\postdate{\end{center}}

\begin{document}

\begin{titlepage}
  \centering
  \vspace*{1cm}
  {\Huge\bfseries Expanding Horizons \\[6pt] \Large Transforming Astronomy in the 2040s \par}
  \vspace{1cm}

  {\LARGE \textbf{GRBs and Relativistic Transients in the 2040s}\par}
  \vspace{1cm}

  \begin{tabular}{p{4.5cm}p{10cm}}
    \textbf{Scientific Categories:} & Time-domain; High-energy astrophysics; Cosmology, Stars \\
    \\
    \textbf{Submitting Author:} & Nikhil Sarin \\
    & Affiliation:  University of Cambridge\\
    & Email: nikhil.sarin@ast.cam.ac.uk \\
    \\
    \textbf{Contributing authors:} & Andrew Levan (Radboud University, a.levan@astro.ru.nl), Nial Tanvir (University of Leicester, nrt3@le.ac.uk), Simone Scaringi (Durham), Fabio Ragosta (University of Naples "Federico II"), Andrea Melandri (INAF. andrea.melandri@inaf.it), Paul Groot (Radboud University, p.groot@astro.ru.nl), Paul O'Brien (University of Leicester, pto2@leicester.ac.uk), Paul Lasky (Monash University, paul.lasky@monash.edu), Samaya Nissanke (DESY, DZA and University of Potsdam, samaya.nissanke@desy.de)
    Alexander Heger (Monash University),
    Steve Schulze (U Northwestern, USA)\\
    \textbf{Endorsers:}  & 
    {
    Igor Andreoni (University of North Carolina, USA), Mattia Bulla (University of Ferrara, IT),
    Teagan Clarke (Princeton University, USA),
    Lise Christensen (University of Copenhagen, DK),
    Ashley Chrimes (ESA/ESTEC, NL),
    Michael Coughlin (UMN, USA),
    Benjamin L. Davis (NYUAD, UAE),
    Valerio D'Elia (Italian Space Agency, IT),
    Dougal Dobie (University of Sydney, Australia),
    Luca Izzo (INAF, Italy),
    Antonio Martin-Carrillo (University College Dublin, IRE),
    Bernhard Müller (Monash University, AU),
    Quentin Pognan (Albert Einstein Institute, DE),
    Benjamin Schneider (Laboratoire d'Astrophysique de Marseille, FR)
    Manisha Shrestha (Monash University, Australia),
    Gokul Srinivasaragavan (University of Maryland College Park, USA)
    Paraskevas Lampropoulos (University of Amsterdam, NL),
    Gianpiero Tagliaferri (INAF, IT),
    Aishwarya Linesh Thakur (INAF, IT)
    Massimiliano De Pasquale (University of Messina, IT), Deanne L. Coppejans (University of Warwick, UK)
    Maria Grazia Bernardini (INAF, Italy)
    Paolo D'Avanzo (INAF, Italy)
    Jonathan Carney (University of North Carolina, USA)
    }
    \\
  \end{tabular}

  \vspace{1cm}

\end{titlepage}

\section{Introduction and Background}
\label{sec:intro}

Relativistic transients---astrophysical phenomena characterized by bulk motion with Lorentz factors $\Gamma \gtrsim 1$ and extreme energy densities---represent the brightest and most energetic explosions in the Universe~\cite{GRBreview}. Since the discovery of gamma-ray bursts in the 1960s, our understanding has expanded to encompass a rich taxonomy, including jetted tidal disruption events (TDEs)~\cite{Bloom2011}, luminous fast blue optical transients (LFBOTs)~\cite{Perley2019}, fast X-ray transients (FXTs)~\cite{Bauer2017}, and ultra-long GRBs~\cite{Levan2014}, spanning the $\gamma$-ray, X-ray, UV/optical/IR, and radio regimes. Each discovery reveals pathways to extreme physics: magnetic field amplification in collisionless shocks~\cite{Sironi2009}, particle acceleration to PeV energies, and relativistic jet launching from compact objects spanning six orders of magnitude in mass. Crucially, these explosions serve dual roles: as laboratories for extreme physics and as powerful cosmic lighthouses. Their exceptional luminosities pierce through dust-obscured regions~\cite{Tanvir2009}, probe the chemical enrichment and gas content of nascent galaxies~\cite{Savaglio2009}, trace stellar evolution endpoints across cosmic time~\cite{Fruchter2006}, and provide unique sightlines for mapping cosmic reionization~\cite{Totani2006}---science inaccessible through any other means.

The 2040s will transform this landscape. Optical surveys such as The Vera C. Rubin Observatory's LSST will discover tens of thousands of optical transients nightly~\cite{lsst}, and \textit{THESEUS} will provide all-sky X-ray/gamma-ray monitoring with $10{-}100\times$ current sensitivity~\cite{THESEUS:2017qvx}. Third-generation gravitational wave observatories (Einstein Telescope, Cosmic Explorer) will detect $\mathcal{O}(10^5)$ compact object mergers annually to $z \sim 2{-}3$, many accompanied by relativistic transients~\cite{Maggiore2020}. High-redshift GRBs will routinely probe $z > 6$, accessing the Universe's first billion years~\cite{Salvaterra2009,Tanvir2012, Ciolfi2021}.

This ``industrial scale'' discovery rate presents unprecedented opportunity. Although the numbers will be large, relativistic transients remain tractable---far rarer than typical transients, enabling homogeneous population studies. Key frontiers include: distinguishing LFBOT engine models through multi-wavelength evolution~\cite{Prentice2018, Perley2019}; mapping black hole demographics via jetted TDEs; identifying successful versus failed jets in supernovae~\cite{Schroeder2025}; and leveraging high-$z$ GRBs as backlights to measure metallicity, dust content, and ionization in galaxies during cosmic dawn and reionization~\cite{Inoue2013}.

Relativistic transients evolve on seconds (prompt emission) to days (afterglow) timescales. Multi-wavelength evolution encodes jet composition, ambient density, magnetic fields, and environments---but only if captured in real time. Photometry traces energetics and geometry; spectroscopy reveals composition and kinematics; polarimetry maps magnetic structure. Current facilities excel individually but struggle with simultaneous rapid-response characterization. Large telescopes provide deep spectroscopy, but cannot tile large regions or track multiple fast-evolving targets; wide-field surveys discover events but lack sensitivity for faint or distant follow-up.

The primary 2040s barrier is not event detection but \textit{immediate, multi-faceted characterization}: (1) rapid localization within large GW localizations or survey alerts; (2) spectroscopic diagnosis of composition, velocities, and environments; (3) high-cadence observations capturing rapid evolution; (4) sufficient depth for cosmological distances. Addressing this requires a paradigm shift: facilities combining temporal agility with observational versatility, managing diverse targets across the sky through adaptive strategies responding to each transient's physical state.

\section{Open Science Questions in the 2040s}
\label{sec:openquestions}

The following open questions represent critical knowledge gaps that will drive relativistic astrophysics in the 2040s. They broadly divide into two components: understanding the extreme physics of relativistic ejecta, and using bright jets as backlights to probe broader astrophysical and cosmological questions.\\
\textbf{2.1 \, How are relativistic jets launched and powered?}: Does prompt gamma-ray emission arise from internal shocks, magnetic reconnection, or photospheric radiation? Time-resolved spectroscopy can distinguish these mechanisms through characteristic spectral evolution. Are the mechanisms powering GRBs, jetted TDEs, and AGN fundamentally the same? And do neutron star central engines launch relativistic jets?\\
\textbf{2.2 \, What is the structure of relativistic outflows?}: Are jets narrow pencil beams or wide structured outflows with angular variations in energy and Lorentz factor? Gravitational wave detections provide viewing angles, enabling jet structure mapping through off-axis observations. Are magnetic fields ordered or tangled, and how does this impact particle acceleration efficiency?\\
\textbf{2.3 \, What determines jet success or failure?}: What fraction of collapsing massive stars launch jets, and what determines whether they successfully break out to produce GRBs versus remaining choked? Do broad-line Type Ic supernovae without detected GRBs always harbor some sort of failed jet? Are FXTs the electromagnetic signature of marginally successful jets, weak GRBs viewed off-axis, or a distinct progenitor population bridging relativistic and non-relativistic transients? How do progenitor mass, metallicity, rotation, and circumstellar environment influence jet breakout probability?\\
\textbf{2.4 \, How do progenitors and environments shape relativistic transients?}: What stars create GRBs and how does this vary across cosmic time? Where in their host galaxies do these events occur, and what are the local star formation rates and metallicities? High-resolution spectroscopy detects narrow metal absorption lines from circumstellar material ejected months to years before explosion, constraining mass-loss histories and binary evolution. At $z > 2{-}3$, these absorption features remain viable probes of the near-progenitor environment, though time-variability studies become increasingly challenging with diminishing flux.\\
\textbf{2.5 \, What sustains long-lived energy injection?}: Why do some GRB afterglows maintain constant luminosity for thousands of seconds---are these powered by rapidly-spinning magnetars or fallback accretion? Do late-time flares represent renewed jet activity or energy propagating through jet structure? Spectroscopy and polarimetry distinguish internal versus external origins.\\
\textbf{2.6 \, What powers the most luminous explosions?}: Are LFBOTs powered by magnetars, fallback accretion onto black holes, or collisions with dense circumstellar shells? How often do massive stars collapse directly to black holes without producing any observable optical transient? What are the observational signatures distinguishing these ultra-faint outcomes from ordinary core collapse?\\
\textbf{2.7 \, What ended the cosmic dark ages?}: Did metal-free Population III stars power the first GRBs? Their afterglows should exhibit pristine chemical signatures---negligible metal absorption and strong He~\textsc{ii} emission---detectable through rapid spectroscopic follow-up of $z > 10$ events. High signal-to-noise spectroscopy of Lyman-$\alpha$ damping wings maps ionized bubble geometry and measures when the Universe became transparent.\\
\textbf{2.8 \, Where were the heavy elements forged?}: Neutron star mergers produce $r$-process elements, but which elements, in what quantities, and with what diversity? Late-time infrared spectroscopy can identify individual heavy elements in kilonova ejecta. With $\sim10^5$ merger detections per year, we can measure how heavy element production varies with redshift, progenitor properties, and host galaxy environment.\\
\textbf{2.9 \, How did early galaxies build up their metals?}: High-redshift GRB spectroscopy measures carbon, oxygen, silicon, and iron abundances in $z > 8$ galaxies, revealing the early chemical enrichment history at the transition from massive Pop~III stars to lower-mass Pop~II stars. Detection of H$_2$, CO, and OH absorption traces molecular gas reservoirs, revealing physical conditions during the first billion years of cosmic history.

\section{Technology Requirements}
\label{sec:tech}

Relativistic transient observations require: (i) trigger mechanisms from high-energy satellites or ground-based rapid variability surveys; (ii) counterpart identification through sensitive optical/IR imaging, particularly for high-$z$ events; (iii) spectroscopic follow-up ($R \sim 2{,}000{-}10{,}000$) for redshifts, energetics, velocity structures, and host properties; (iv) long-term multi-wavelength monitoring (optical/IR/X-ray/radio) mapping blast wave evolution. These transients span extreme dynamic range---brightest afterglows reach 10$^{\mathrm{th}}$ magnitude whereas faint distant events may be $R > 25$---requiring flexible 1--10+ m aperture access.

Time-critical observations demand rapid, flexible scheduling and comprehensive real-time pipelines delivering science-grade products. ESO has substantial expertise here, and relativistic transient requirements align closely with existing optical/IR capabilities. Maximizing scientific return, however, requires enhanced robotic scheduling enabling dynamic, autonomous observation decisions. Future improvements should include automatic counterpart identification, spectral acquisition, and redshift measurement, minimizing latency for rapid characterization.

\textbf{Required capabilities:} A large aperture ($10{-}30\,$m effective collecting area), fully flexible/scalable optical-NIR spectroscopic facility capable of targeting 5--10+ events per night down to $m \sim 23{-}25$, with the ability to perform time-critical observations at short notice (minutes to hours). Spectrographs covering optical to NIR wavelengths at $R \sim 2{,}000{-}10{,}000$ are essential for identifying heavy element lines, measuring velocity structures, and detecting high-$z$ absorption features. Fast response, i.e., slewing and locking within seconds, is critical for capturing shock breakouts, pre-merger precursors, and early-time emission encoding jet physics. Multi-wavelength capabilities with comparable sensitivity (e.g. sub-mm, radio) are also valuable for jet-physics.

In addition, small field-of-view ($\sim$ few arcmin) imaging capability is valuable for counterpart identification within moderate error boxes (e.g., Swift-like localizations), complementing wide-field survey discoveries. This capability, achievable through acquisition cameras offset from fiber feeds, bridges the gap between large localization regions and spectroscopic characterization.

\textbf{Complementarity with ESO facilities:} Current ESO plans for the 2040s (VLT continuation, ELT operations, ALMA) do not include a dedicated, robotically-operated time-domain facility optimized for rapid-response spectroscopy across multiple simultaneous targets. The VLT and ELT, although providing deep spectroscopic capability, operate on classical/queue scheduling incompatible with the rapid, flexible, multi-target demands of industrial-scale transient follow-up. A dedicated time-domain facility would complement these capabilities, enabling ESO to fully exploit the 2040s transient landscape while preserving traditional facility operations for their primary science cases.

\printbibliography

@article{Ciolfi2021,
   title={Multi-messenger astrophysics with THESEUS in the 2030s},
   volume={52},
   ISSN={1572-9508},
   url={http://dx.doi.org/10.1007/s10686-021-09795-9},
   DOI={10.1007/s10686-021-09795-9},
   number={3},
   journal={Experimental Astronomy},
   publisher={Springer Science and Business Media LLC},
   author={Ciolfi, Riccardo and Stratta, Giulia and Branchesi, Marica and Gendre, Bruce and Grimm, Stefan and Harms, Jan and Lamb, Gavin Paul and Martin-Carrillo, Antonio and McCann, Ayden and Oganesyan, Gor and Palazzi, Eliana and Ronchini, Samuele and Rossi, Andrea and Salafia, Om Sharan and Salmon, Lana and Ascenzi, Stefano and Capone, Antonio and Celli, Silvia and Dall’Osso, Simone and Di Palma, Irene and Fasano, Michela and Fermani, Paolo and Guetta, Dafne and Hanlon, Lorraine and Howell, Eric and Paltani, Stephane and Rezzolla, Luciano and Vinciguerra, Serena and Zegarelli, Angela and Amati, Lorenzo and Blain, Andrew and Bozzo, Enrico and Chaty, Sylvain and D’Avanzo, Paolo and De Pasquale, fnmMassimiliano and Dereli-Bégué, Hüsne and Ghirlanda, Giancarlo and Gomboc, Andreja and Götz, Diego and Horvath, Istvan and Hudec, Rene and Izzo, Luca and Le Floch, Emeric and Li, Liang and Longo, Francesco and Komossa, S. and Kong, Albert K. H. and Mereghetti, Sandro and Mignani, Roberto and Nathanail, Antonios and O’Brien, Paul T. and Osborne, Julian P. and Pe’er, Asaf and Piranomonte, Silvia and Rosati, Piero and Savaglio, Sandra and Schüssler, Fabian and Sergijenko, Olga and Shao, Lijing and Tanvir, Nial and Turriziani, Sara and Urata, Yuji and van Putten, Maurice and Vergani, Susanna and Zane, Silvia and Zhang, Bing},
   year={2021},
   month=oct, pages={245–275} }

@article{Maggiore2020,
  title = {Science Case for the {{Einstein}} Telescope},
  author = {Maggiore, Michele and Broeck, Chris Van Den and Bartolo, Nicola and Belgacem, Enis and Bertacca, Daniele and Bizouard, Marie Anne and Branchesi, Marica and Clesse, Sebastien and Foffa, Stefano and {Garc{\'i}a-Bellido}, Juan and Grimm, Stefan and Harms, Jan and Hinderer, Tanja and Matarrese, Sabino and Palomba, Cristiano and Peloso, Marco and Ricciardone, Angelo and Sakellariadou, Mairi},
  year = 2020,
  month = mar,
  journal = {Journal of Cosmology and Astroparticle Physics},
  volume = {2020},
  number = {03},
  pages = {050},
  doi = {10.1088/1475-7516/2020/03/050}
}

@ARTICLE{lsst,
       author = {{Ivezi{\'c}}, {\v{Z}}. and {Kahn}, S. M. and {Tyson}, J. Anthony and {Abel}, Bob and {Acosta}, Emily and {Allsman}, Robyn and {Alonso}, David and {AlSayyad}, Yusra and {Anderson}, Scott F. and {Andrew}, John and {Angel}, James Roger P. and {Angeli}, George Z. and {Ansari}, Reza and {Antilogus}, Pierre and {Araujo}, Constanza and {Armstrong}, Robert and {Arndt}, Kirk T. and {Astier}, Pierre and {Aubourg}, {\'E}ric and {Auza}, Nicole and {Axelrod}, Tim S. and {Bard}, Deborah J. and {Barr}, Jeff D. and {Barrau}, Aurelian and {Bartlett}, James G. and {Bauer}, Amanda E. and {Bauman}, Brian J. and {Baumont}, Sylvain and {Bechtol}, Ellen and {Bechtol}, Keith and {Becker}, Andrew C. and {Becla}, Jacek and {Beldica}, Cristina and {Bellavia}, Steve and {Bianco}, Federica B. and {Biswas}, Rahul and {Blanc}, Guillaume and {Blazek}, Jonathan and {Blandford}, Roger D. and {Bloom}, Josh S. and {Bogart}, Joanne and {Bond}, Tim W. and {Booth}, Michael T. and {Borgland}, Anders W. and {Borne}, Kirk and {Bosch}, James F. and {Boutigny}, Dominique and {Brackett}, Craig A. and {Bradshaw}, Andrew and {Brandt}, William Nielsen and {Brown}, Michael E. and {Bullock}, James S. and {Burchat}, Patricia and {Burke}, David L. and {Cagnoli}, Gianpietro and {Calabrese}, Daniel and {Callahan}, Shawn and {Callen}, Alice L. and {Carlin}, Jeffrey L. and {Carlson}, Erin L. and {Chandrasekharan}, Srinivasan and {Charles-Emerson}, Glenaver and {Chesley}, Steve and {Cheu}, Elliott C. and {Chiang}, Hsin-Fang and {Chiang}, James and {Chirino}, Carol and {Chow}, Derek and {Ciardi}, David R. and {Claver}, Charles F. and {Cohen-Tanugi}, Johann and {Cockrum}, Joseph J. and {Coles}, Rebecca and {Connolly}, Andrew J. and {Cook}, Kem H. and {Cooray}, Asantha and {Covey}, Kevin R. and {Cribbs}, Chris and {Cui}, Wei and {Cutri}, Roc and {Daly}, Philip N. and {Daniel}, Scott F. and {Daruich}, Felipe and {Daubard}, Guillaume and {Daues}, Greg and {Dawson}, William and {Delgado}, Francisco and {Dellapenna}, Alfred and {de Peyster}, Robert and {de Val-Borro}, Miguel and {Digel}, Seth W. and {Doherty}, Peter and {Dubois}, Richard and {Dubois-Felsmann}, Gregory P. and {Durech}, Josef and {Economou}, Frossie and {Eifler}, Tim and {Eracleous}, Michael and {Emmons}, Benjamin L. and {Fausti Neto}, Angelo and {Ferguson}, Henry and {Figueroa}, Enrique and {Fisher-Levine}, Merlin and {Focke}, Warren and {Foss}, Michael D. and {Frank}, James and {Freemon}, Michael D. and {Gangler}, Emmanuel and {Gawiser}, Eric and {Geary}, John C. and {Gee}, Perry and {Geha}, Marla and {Gessner}, Charles J.~B. and {Gibson}, Robert R. and {Gilmore}, D. Kirk and {Glanzman}, Thomas and {Glick}, William and {Goldina}, Tatiana and {Goldstein}, Daniel A. and {Goodenow}, Iain and {Graham}, Melissa L. and {Gressler}, William J. and {Gris}, Philippe and {Guy}, Leanne P. and {Guyonnet}, Augustin and {Haller}, Gunther and {Harris}, Ron and {Hascall}, Patrick A. and {Haupt}, Justine and {Hernandez}, Fabio and {Herrmann}, Sven and {Hileman}, Edward and {Hoblitt}, Joshua and {Hodgson}, John A. and {Hogan}, Craig and {Howard}, James D. and {Huang}, Dajun and {Huffer}, Michael E. and {Ingraham}, Patrick and {Innes}, Walter R. and {Jacoby}, Suzanne H. and {Jain}, Bhuvnesh and {Jammes}, Fabrice and {Jee}, M. James and {Jenness}, Tim and {Jernigan}, Garrett and {Jevremovi{\'c}}, Darko and {Johns}, Kenneth and {Johnson}, Anthony S. and {Johnson}, Margaret W.~G. and {Jones}, R. Lynne and {Juramy-Gilles}, Claire and {Juri{\'c}}, Mario and {Kalirai}, Jason S. and {Kallivayalil}, Nitya J. and {Kalmbach}, Bryce and {Kantor}, Jeffrey P. and {Karst}, Pierre and {Kasliwal}, Mansi M. and {Kelly}, Heather and {Kessler}, Richard and {Kinnison}, Veronica and {Kirkby}, David and {Knox}, Lloyd and {Kotov}, Ivan V. and {Krabbendam}, Victor L. and {Krughoff}, K. Simon and {Kub{\'a}nek}, Petr and {Kuczewski}, John and {Kulkarni}, Shri and {Ku}, John and {Kurita}, Nadine R. and {Lage}, Craig S. and {Lambert}, Ron and {Lange}, Travis and {Langton}, J. Brian and {Le Guillou}, Laurent and {Levine}, Deborah and {Liang}, Ming and {Lim}, Kian-Tat and {Lintott}, Chris J. and {Long}, Kevin E. and {Lopez}, Margaux and {Lotz}, Paul J. and {Lupton}, Robert H. and {Lust}, Nate B. and {MacArthur}, Lauren A. and {Mahabal}, Ashish and {Mandelbaum}, Rachel and {Markiewicz}, Thomas W. and {Marsh}, Darren S. and {Marshall}, Philip J. and {Marshall}, Stuart and {May}, Morgan and {McKercher}, Robert and {McQueen}, Michelle and {Meyers}, Joshua and {Migliore}, Myriam and {Miller}, Michelle and {Mills}, David J. and {Miraval}, Connor and {Moeyens}, Joachim and {Moolekamp}, Fred E. and {Monet}, David G. and {Moniez}, Marc and {Monkewitz}, Serge and {Montgomery}, Christopher and {Morrison}, Christopher B. and {Mueller}, Fritz and {Muller}, Gary P. and {Mu{\~n}oz Arancibia}, Freddy and {Neill}, Douglas R. and {Newbry}, Scott P. and {Nief}, Jean-Yves and {Nomerotski}, Andrei and {Nordby}, Martin and {O'Connor}, Paul and {Oliver}, John and {Olivier}, Scot S. and {Olsen}, Knut and {O'Mullane}, William and {Ortiz}, Sandra and {Osier}, Shawn and {Owen}, Russell E. and {Pain}, Reynald and {Palecek}, Paul E. and {Parejko}, John K. and {Parsons}, James B. and {Pease}, Nathan M. and {Peterson}, J. Matt and {Peterson}, John R. and {Petravick}, Donald L. and {Libby Petrick}, M.~E. and {Petry}, Cathy E. and {Pierfederici}, Francesco and {Pietrowicz}, Stephen and {Pike}, Rob and {Pinto}, Philip A. and {Plante}, Raymond and {Plate}, Stephen and {Plutchak}, Joel P. and {Price}, Paul A. and {Prouza}, Michael and {Radeka}, Veljko and {Rajagopal}, Jayadev and {Rasmussen}, Andrew P. and {Regnault}, Nicolas and {Reil}, Kevin A. and {Reiss}, David J. and {Reuter}, Michael A. and {Ridgway}, Stephen T. and {Riot}, Vincent J. and {Ritz}, Steve and {Robinson}, Sean and {Roby}, William and {Roodman}, Aaron and {Rosing}, Wayne and {Roucelle}, Cecille and {Rumore}, Matthew R. and {Russo}, Stefano and {Saha}, Abhijit and {Sassolas}, Benoit and {Schalk}, Terry L. and {Schellart}, Pim and {Schindler}, Rafe H. and {Schmidt}, Samuel and {Schneider}, Donald P. and {Schneider}, Michael D. and {Schoening}, William and {Schumacher}, German and {Schwamb}, Megan E. and {Sebag}, Jacques and {Selvy}, Brian and {Sembroski}, Glenn H. and {Seppala}, Lynn G. and {Serio}, Andrew and {Serrano}, Eduardo and {Shaw}, Richard A. and {Shipsey}, Ian and {Sick}, Jonathan and {Silvestri}, Nicole and {Slater}, Colin T. and {Smith}, J. Allyn and {Smith}, R. Chris and {Sobhani}, Shahram and {Soldahl}, Christine and {Storrie-Lombardi}, Lisa and {Stover}, Edward and {Strauss}, Michael A. and {Street}, Rachel A. and {Stubbs}, Christopher W. and {Sullivan}, Ian S. and {Sweeney}, Donald and {Swinbank}, John D. and {Szalay}, Alexander and {Takacs}, Peter and {Tether}, Stephen A. and {Thaler}, Jon J. and {Thayer}, John Gregg and {Thomas}, Sandrine and {Thornton}, Adam J. and {Thukral}, Vaikunth and {Tice}, Jeffrey and {Trilling}, David E. and {Turri}, Max and {Van Berg}, Richard and {Vanden Berk}, Daniel and {Vetter}, Kurt and {Virieux}, Francoise and {Vucina}, Tomislav and {Wahl}, William and {Walkowicz}, Lucianne and {Walsh}, Brian and {Walter}, Christopher W. and {Wang}, Daniel L. and {Wang}, Shin-Yawn and {Warner}, Michael and {Wiecha}, Oliver and {Willman}, Beth and {Winters}, Scott E. and {Wittman}, David and {Wolff}, Sidney C. and {Wood-Vasey}, W. Michael and {Wu}, Xiuqin and {Xin}, Bo and {Yoachim}, Peter and {Zhan}, Hu},
       title = "{LSST: From Science Drivers to Reference Design and Anticipated Data Products}",
      journal = {APJ},
       year = 2019,
       month = mar,
       volume = {873},
       number = {2},
       eid = {111},
       pages = {111},
       doi = {10.3847/1538-4357/ab042c},
archivePrefix = {arXiv},
       eprint = {0805.2366},
 primaryClass = {astro-ph},
       adsurl = {https://ui.adsabs.harvard.edu/abs/2019ApJ...873..111I}
}

@article{GRBreview,
       author = {{Kumar}, P. and {Zhang}, B.},
        title = "{The physics of gamma-ray bursts \& relativistic jets}",
      journal = {\physrep},
         year = 2015,
        month = feb,
       volume = {561},
        pages = {1-109},
          doi = {10.1016/j.physrep.2014.09.008},
       adsurl = {https://ui.adsabs.harvard.edu/abs/2015PhR...561....1K}
}

@article{THESEUS:2017qvx,
    author = "Amati, L. and others",
    collaboration = "THESEUS",
    title = "{The THESEUS space mission concept: science case, design and expected performances}",
    eprint = "1710.04638",
    archivePrefix = "arXiv",
    primaryClass = "astro-ph.IM",
    doi = "10.1016/j.asr.2018.03.010",
    journal = "{Adv. Space Res.}",
    volume = "62",
    pages = "191--244",
    year = "2018"
}

@ARTICLE{Bloom2011,
       author = {{Bloom}, Joshua S. and {Giannios}, Dimitrios and {Metzger}, Brian D. and {Cenko}, S. Bradley and et al.},
        title = "{A Possible Relativistic Jetted Outburst from a Massive Black Hole Fed by a Tidally Disrupted Star}",
      journal = {Science},
         year = 2011,
        month = jul,
       volume = {333},
       number = {6039},
        pages = {203},
          doi = {10.1126/science.1207150},
archivePrefix = {arXiv},
       eprint = {1104.3257},
 primaryClass = {astro-ph.HE},
       adsurl = {https://ui.adsabs.harvard.edu/abs/2011Sci...333..203B}
}

@ARTICLE{Perley2019,
       author = {{Perley}, Daniel A. and {Mazzali}, Paolo A. and {Yan}, Lin and {Cenko}, S. Bradley and et al.},
        title = "{The fast, luminous ultraviolet transient AT2018cow: extreme supernova, or disruption of a star by an intermediate-mass black hole?}",
      journal = {\mnras},
         year = 2019,
        month = mar,
       volume = {484},
       number = {1},
        pages = {1031-1049},
          doi = {10.1093/mnras/sty3420},
archivePrefix = {arXiv},
       eprint = {1808.00969},
 primaryClass = {astro-ph.HE},
       adsurl = {https://ui.adsabs.harvard.edu/abs/2019MNRAS.484.1031P}
}

@ARTICLE{Bauer2017,
       author = {{Bauer}, Franz E. and {Treister}, Ezequiel and {Schawinski}, Kevin and {Schulze}, Steve and et al.},
        title = "{A new, faint population of X-ray transients}",
      journal = {\mnras},
         year = 2017,
        month = jun,
       volume = {467},
       number = {4},
        pages = {4841-4857},
          doi = {10.1093/mnras/stx417},
archivePrefix = {arXiv},
       eprint = {1702.04422},
 primaryClass = {astro-ph.HE},
       adsurl = {https://ui.adsabs.harvard.edu/abs/2017MNRAS.467.4841B}
}

@ARTICLE{Levan2014,
       author = {{Levan}, A.~J. and {Tanvir}, N.~R. and {Starling}, R.~L.~C. and {Wiersema}, K. and et al.},
        title = "{A New Population of Ultra-long Duration Gamma-Ray Bursts}",
      journal = {\apj},
         year = 2014,
        month = jan,
       volume = {781},
       number = {1},
          eid = {13},
        pages = {13},
          doi = {10.1088/0004-637X/781/1/13},
archivePrefix = {arXiv},
       eprint = {1302.2352},
 primaryClass = {astro-ph.HE},
       adsurl = {https://ui.adsabs.harvard.edu/abs/2014ApJ...781...13L}
}

@ARTICLE{Sironi2009,
       author = {{Sironi}, Lorenzo and {Spitkovsky}, Anatoly},
        title = "{Particle Acceleration in Relativistic Magnetized Collisionless Pair Shocks: Dependence of Shock Acceleration on Magnetic Obliquity}",
      journal = {\apj},
         year = 2009,
        month = jun,
       volume = {698},
       number = {2},
        pages = {1523-1549},
          doi = {10.1088/0004-637X/698/2/1523},
archivePrefix = {arXiv},
       eprint = {0901.2578},
 primaryClass = {astro-ph.HE},
       adsurl = {https://ui.adsabs.harvard.edu/abs/2009ApJ...698.1523S}
}

@ARTICLE{Tanvir2009,
       author = {{Tanvir}, N.~R. and {Fox}, D.~B. and {Levan}, A.~J. and {Berger}, E. and et al.},
        title = "{A {\ensuremath{\gamma}}-ray burst at a redshift of z\raisebox{-0.5ex}\textasciitilde8.2}",
      journal = {\nat},
         year = 2009,
        month = oct,
       volume = {461},
       number = {7268},
        pages = {1254-1257},
          doi = {10.1038/nature08459},
archivePrefix = {arXiv},
       eprint = {0906.1577},
 primaryClass = {astro-ph.CO},
       adsurl = {https://ui.adsabs.harvard.edu/abs/2009Natur.461.1254T}
}

@ARTICLE{Savaglio2009,
       author = {{Savaglio}, S. and {Glazebrook}, K. and {Le Borgne}, D.},
        title = "{The Galaxy Population Hosting Gamma-Ray Bursts}",
      journal = {\apj},
         year = 2009,
        month = jan,
       volume = {691},
       number = {1},
        pages = {182-211},
          doi = {10.1088/0004-637X/691/1/182},
archivePrefix = {arXiv},
       eprint = {0803.2718},
 primaryClass = {astro-ph},
       adsurl = {https://ui.adsabs.harvard.edu/abs/2009ApJ...691..182S}
}

@ARTICLE{Fruchter2006,
       author = {{Fruchter}, A.~S. and {Levan}, A.~J. and {Strolger}, L. and {Vreeswijk}, P.~M. and et al.},
        title = "{Long {\ensuremath{\gamma}}-ray bursts and core-collapse supernovae have different environments}",
      journal = {\nat},
         year = 2006,
        month = may,
       volume = {441},
       number = {7092},
        pages = {463-468},
          doi = {10.1038/nature04787},
archivePrefix = {arXiv},
       eprint = {astro-ph/0603537},
 primaryClass = {astro-ph},
       adsurl = {https://ui.adsabs.harvard.edu/abs/2006Natur.441..463F}
}

@ARTICLE{Totani2006,
       author = {{Totani}, Tomonori and {Kawai}, Nobuyuki and {Kosugi}, George and {Aoki}, Kentaro and et al.},
        title = "{Implications for Cosmic Reionization from the Optical Afterglow Spectrum of the Gamma-Ray Burst 050904 at z = 6.3$^{*}$}",
      journal = {\pasj},
         year = 2006,
        month = jun,
       volume = {58},
       number = {3},
        pages = {485-498},
          doi = {10.1093/pasj/58.3.485},
archivePrefix = {arXiv},
       eprint = {astro-ph/0512154},
 primaryClass = {astro-ph},
       adsurl = {https://ui.adsabs.harvard.edu/abs/2006PASJ...58..485T}
}

\end{document}